\documentstyle[11pt,emulateapj,psfig]{article}
\begin{document}



\title{A note on the discovery of absorption features in 1E 1207.4-5209$^{1}$}

\author{R.X. Xu, H.G. Wang, G.J. Qiao\\
        School of Physics, Peking University, Beijing 100871,
        China}


\altaffiltext{1}{This work is supported by NSFC (No. 10173002),
and by the Special Funds for Major State Basic Research Projects
of China (G2000077602).}

\begin{abstract}

In a paper by Sanwal et al. (2002), it is supposed to be very
difficult to interpret the absorption features in term of
cyclotron lines. However, we would like to address here that the
possibility of the absorption being cyclotron resonance can not be
ruled out.
We propose that the isolate neutron star, 1E 1207.4-5209 in the
center of supernova remnant PKS 1209-51/52, has a debris disk and
is in a propeller phase, with an accretion rate $\sim 6\times
10^{-11}M_\odot$/year.
In this scenario, 1E 1207.4-5209 could also be a bare strange
star.

\vspace{0.4cm} %
\noindent %

\end{abstract}



\section{Introduction}

Strange (quark) star is composed of nearly equal number of up,
down, and strange quarks, and a few electrons for keeping
neutralization of matter.
It has important implications for studying the phase diagram of
strong interaction system whether this kind of quark stars exist.
Recently, Xu (2002) suggests that a featureless thermal spectrum
could be a probe for identifying strange stars, since no bound
charged particle is in discrete quantum states on the quark
surface without strong magnetic field.
Nonetheless, it is worth noting that discrete Landau levels appear
for charged particles in strong fields.

Two absorption lines, at $\sim 0.7$ and $\sim 1.4$ keV, are
detected from an isolate neutron star (1E 1207.4-5209) with {\it
Chandra} by Sanwal et al. (2002), and are then confirmed with {\it
XMM-Newton} by Mereghetti et al. (2002).
Certainly 1E 1207.4-5209 can not be a bare strange star if those
two lines are atomic-transition originated, although the stellar
mass $M$ and radius $R$ may be derived by obtaining the
gravitational redshift (as $M/R$) and the pressure broadening (as
$M/R^2$) of the lines.
However, if these double lines are caused by the Landau-level
transition of electrons, 1E 1207.4-5209 could also be a bare
strange star since no atom might be on the stellar surface.
Sanwal et al. (2002) addressed that the features are associated
with atomic transition of once-ionized helium, and thought that it
is hard to interpret the absorption features in term of cyclotron
lines.
However, we will find in the next section that the possibility of
the absorption being cyclotron resonance can not be ruled out.
We will present, in this {\em Letter}, a short note to confute the
interpretation of the recently discovered lines in the X-ray
spectrum of 1E1207-52, that these spectral lines are testifying
the presence of an atmosphere on the star which can absolutely not
be a bare strange star.

\section{The possibility of cyclotron lines from 1E 1207.4-5209}

In Sanwal et al. (2002), the authors discussed two potential
possibilities in generating the absorption features, i.e.,
cyclotron or atomic transition lines, but considered the former
scenario unlikely. They hence suggested that the features are
associated with atomic transitions of once-ionized helium. Here we
hope to point out that all the criticisms about the cyclotron line
mechanism by Sanwal et al. can be circumvented, so that such a
possibility is not ruled out. Below we will answer their
criticisms in turn.

1. The inferred field value, $B_{\rm e}=3\times 10^{12}$ G, based
on $P$ and $\dot P$ is significantly larger than the field $B_{\rm
ce}\sim 10^{11}$ G derived by assuming that the 0.7 keV and
$0.7\times 2$ keV lines are the fundamental and the first
harmonics, respectively (Sanwal et al. 2002).
However, it is possible that 1E 1207.4-5209 has a debris disk,
which is currently conjectured for interpreting the enigmatic
sources of anomalous X-ray pulsars and soft $\gamma-$ray repeaters
in literatures (e.g., Chatterjee et al. 2000), since it is in the
center of a supernova remnant.
A recent discovery of a near-infrared counterpart to an AXP
(1E1048, Wang \& Chakrabarty 2002) may be a hint of such kind a
fallback accretion disk.
Israel et al.'s discovery (2002) strengthens the infrared
association with the AXP 1E1048.1-5937.
In case of this disk, 1E 1207.4-5209 provides a potential
possibility for us to investigate the disk property by dividing
the total braking rate into magnetodipole and disk ones, since the
total rotational energy loss ($I\Omega{\dot \Omega}$, $I$ the
moment of inertia) is the sum of the dipole radiation power and
the propeller energy loss ${\dot E}_{\rm d}=I\Omega{\dot
\Omega}_{\rm d}\simeq -G{\dot M}M/R_{\rm m}$ (Francischelli \&
Wijers 2002), where $M$ is the mass of center star, $\dot M$ the
accretion rate, and the magnetosphere radius $R_{\rm m}\simeq
2.2\times 10^{13}B_{12}^{4/7}{\dot M}^{-2/7}$ cm.
It is worth noting that most of the accretion material driven by
viscosity in the fossil disk can't fall onto the stellar surface
because of centrifugal forces acting on the matter, but is ejected
by the boundary reaction during the propeller phase.
The magnetodipole spin down ${\dot \Omega}_{\rm m}\simeq -8\times
10^{-16}$ s$^{-2}$ for a pulsar with polar field $B_{\rm ce}$, and
then ${\dot \Omega}_{\rm d}={\dot \Omega}-{\dot \Omega}_{\rm
m}\simeq -7\times 10^{-13}$ s$^{-2}$.
The fossil disk torque dominates, which could be the reason that
this source is radio quiet (otherwise, a source may be radio loud
if the dipole radiation torque dominates).
Therefore, the accretion rate ${\dot M}\simeq 9.1\times 10^{15}$
g/s $\sim 6.4 \times 10^{-11}M_{\odot}$ per year.
We find $R_{\rm m}\simeq 1.6\times 10^8$ cm, the radius of light
cylinder $R_{\rm L}\simeq 2\times 10^9$ cm, and the corotating
radius $R_{\rm C}\simeq 9.5\times 10^7$ cm. These radii are
consistent with the propeller requirement: $R_{\rm C}\la R_{\rm
m}<R_{\rm L}$.
In addition, the age problem (e.g., Mereghetti et al. 2002), that
the characteristic age $\tau_{\rm c}=200-900$ kyrs is much larger
than the estimated age $\sim 7$ kyrs for the remnant PKS
1209-51/52, might also be solved in this scenario.

2. The 0.7 keV line is not much stronger than the 1.4 keV one
(Sanwal et al. 2002).
It is true that the oscillator strength of the first harmonic is
much smaller than that of the fundamental in the weak field limit,
but this does not mean that the absorption-like dips would have
significant differences. In fact, the spectrum profile should be
calculated by modelling the resonant cyclotron radiation transfer.
For instance, even for a field of $B=1.7\times 10^{12}$ G (in this
case, the ratio of the oscillator strength could be $\sim 0.04$),
those two spectrum lines calculated could be similar in depth
(Freeman et al. 1999), depending on radiative geometry.
Observationally, the cyclotron absorption depth of the fundamental
is {\em not} much stronger than that of the first harmonic (e.g.,
Tr\"umper et al. 1978, with $B\sim 3\times 10^{12}$ G).
In addition, the observation does show that the integrated photons
absorbed by the fundamental transition are much more than those by
the first harmonic transition.

3. The charge density in the star's magnetosphere can not be large
enough to scatter resonantly the photons from the surface (Sanwal
et al. 2002).
The main reason, which leads the authors to this conclusion, is
that the required electron number density, $n_{\rm e}\sim 10^{13}$
cm$^{-3}$, is two orders larger than the Goldreich-Julian density,
$n_{\rm GJ}\sim 5\times 10^{11}$ cm$^{-3}$.
However, although pulsar magnetospheres are not known with
certain, it is a common point that primary pairs with Lorentz
factor $\gamma_{\rm p}\sim 10^6$ and with density $\sim n_{\rm
GJ}$ are accelerated in gaps while more secondary pairs with
Lorentz factor $\gamma_{\rm s}\sim 10^{2-4}$ are created outside
the gaps (e.g., Ruderman \& Sutherland 1975).
In despite of that the net charge density could be $n_{\rm GJ}$,
the absolute number density should be $\sim 10^{2-4}$ times of
$n_{\rm GJ}$, which should be enough to effectively scatter the
photons from the stellar surface.
Actually, this could be another possibility for cyclotron
absorptions in the magnetosphere even if no fossil disk
contributes a braking torque.

4. The atomic transition of once-ionized helium may be responsible
to these lines, although the authors had not presented a full
discussion in this possibility.
However, in this interpretation, they assume a general field with
superstrong strength, $B\simeq 1.5\times 10^{14}$ G, which is much
larger than the derived field $B_{\rm e}=3\times 10^{12}$ G.
It is hard to understand that this ``neutron'' star with a {\em
typical} field $B_{\rm e}$ can have such a strong prevalent
multipole field on the surface.

\section{Discussions}

Certainly Sanwal et al's discovery is very important in both
possibility: the mass and radius may be derived if the absorption
are atomic transition originated, or the accretion rate in the
propeller phase could be estimated, for the first time, in case of
two cyclotron lines (point 1).

Mereghetti et al. (2002) found that the absorption features are
phase-dependent: the $\sim 1.4$ keV line prefers to appear during
the minimum and the rising the parts, rather than at the peak, of
the pulse profile.
This observational property may reflect the geometry of resonant
cyclotron emission: there is almost only one fundamental line for
an observer along the magnetic fields, while more harmonic lines
appear if the line-of-sight is perpendicular to the fields (Fig.3
of Freeman et al. 1999).
An effort to fit the observed spectrum of 1E 1207.4-5209 was tried
by Hailey \& Mori (2002) who presumed that the star has an
atmosphere with He-like Oxygen or Neon in not too high a field;
whereas an elaborate model calculation (being prepared), to fit in
term of cyclotron resonance lines, is also necessary in order to
know the details of the source.

A very interesting and important question is: why is 1E
1207.4-5209 the only one in which significant absorption features
have been detected so far?
To answer this question, Mereghetti et al. (2002) suggested 1E
1207.4-5209 has a metal atmosphere, which is not old enough to
accrete a hydrogen layer.
However, this question may naturally be answered by the selective
effect in observations, since maybe only a few sources have
magnetic fields being suitable for creating cyclotron lines with
energies in the detector energy range.
The fundamental electron cyclotron resonance lies at $\Delta E =
11.6 B_{12}\sqrt{1-R_{\rm s}/R}$ keV, where $B_{12}$ is the polar
magnetic field in $10^{12}$ G, $R_{\rm s}\equiv 2GM/c^2$ is the
Schwarzschild radius, and $M$ and $R$ are the stellar mass and
radius, respectively. For a bare strange star with certain mass
$M$, one can obtain its radius $R$ by integrating numerically the
TOV equation, with the inclusion of the equation of state for
strange matter: $P=(\rho-4B)/3$ ($B$ is the bag constant). The
fundamental resonance energy as a function of magnetic field for
strange stars with different masses is shown in Fig.1.
We can see that, for detectors ({\it Chandra} or {\it XMM-Newton})
from $\sim 0.1$ to $\sim 10$ keV, the sensitivity fields in which
electrons can absorb resonantly photons within that energy range
are from $9\times 10^9$ G to $1\times 10^{12}$G.
It is well known that pulsars tend to have a magnetic field of
$\sim 10^{12}$ G (normal pulsars) or of $\sim 10^8$ G (millisecond
pulsars); it is thus not a surprise that only few sources are
observed to show spectral lines.
{\em No source} listed in the table of Xu (2002) has definitely a
suitable field.

Recently, a 5 keV absorption feature has been detected and
confirmed in the bursts of a soft-gamma-repeater SGR 1806-20
(Ibrahim et al. 2002a, 2002b), which is believed to be the feature
as one of the proton cyclotron lines in superstrong magnetic field
($\sim 10^{15}$ G) by the authors.
However there are some difficulties in this explanation. 1, due to
the high mass-energy ($\sim 1$ GeV) of a proton, the ratio of the
oscillator strength of the first harmonic to that of fundamental
in $10^{15}$ G is {\em only} $\sim 10^{-6}$! It is not reasonable
to detect the first and the {\em even} higher harmonics. In fact,
numerical spectrum simulations of atmospheres with protons in
superstrong fields have never show more than two proton absorption
lines (Ho \& Lai 2001, 2002). 2, a better and more reasonable
model for the continuum spectrum component is needed in order to
identify such absorption features in reality.
Motivated by these flaws, we suggest that the possible absorption
lines at $\sim 5$, $\sim 11.2$, and $\sim 17.5$ keV could be
interpreted as electron cyclotron lines, while the $\sim 7.5$ keV
absorption might be caused by other effects (e.g., can the
accreting plasma with irons absorb at $\sim 7.5$ keV?).
The much small ratio, $\sim 10^{-7}$, of oscillation strength can
also not large enough to produce a second harmonic of $\alpha$
particle at $\sim 7.5$ keV.
SGR 1806-20 may have an ordinary magnetic field, $\sim 5\times
10^{11}$ G, which should be another pulsar-like compact stars with
suitable magnetic fields for the detectors in the sky.

Strange stars could exist; the exotic surface of a bare strange
star might eventually result in the identification of them,
especially the most probable one RX J1856 (e.g., Drake et al.
2002, Xu 2002).
Although each of the observed phenomena from pulsar-like stars may
be interpreted under the regime of traditional neutron star with
unusual or artificial physical properties, it might be a natural
way to understand the observations by updating ``neutron'' stars
with (bare) strange stars.

{\it Acknowledgments.}~~
We would like to thank Dr. Bing Zhang for his valuable
suggestions. RXX wishes to thank Dr. Jianrong Shi for his valuable
discussions about cyclotron line formation.

\clearpage
\begin{figure}
\centerline{\psfig{figure=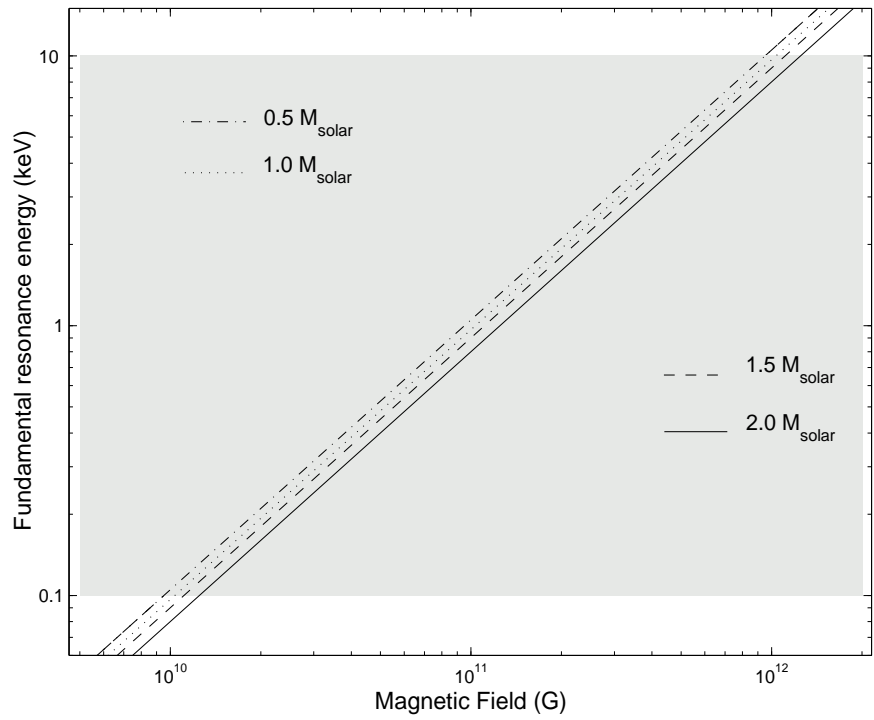,angle=0,height=10cm,width=12cm}}
\caption[]{%
The energy of the fundamental resonance cyclotron as a function of
stellar magnetic fields for bare strange stars with masses of
0.5$M_\odot$, 1.0$M_\odot$, 1.5$M_\odot$, and 2.0$M_\odot$,
respectively, from up to down. The hatched region is from 0.1 keV
to 10 keV, photons with energy within which can be collected
effectively in {\em Chandra} and {\em XMM-Newton} detectors.
}%
\end{figure}


\vspace{1cm}

\begin{thebibliography}{}

\bibitem[]{} Chatterjee, P., Hernquist, L., Narayan, R. 2000, ApJ, 534, 373

\bibitem[]{} Drake, J.J., Marshall, H.L., Dreizler, S., et al.,
2002, ApJ, 572, 996

\bibitem[]{} Francischelli, G. J., Wijers, R. A. M. J. 2002, ApJ,
submitted (astro-ph/0205212)


\bibitem[]{} Freeman, P. E., et al. 1999, ApJ, 524, 772

\bibitem[]{} Hailey, C.J., Mori, K. 2002, ApJL, in press
(astro-ph/0207590)

\bibitem[]{} Ho, W. C. G., Lai, D. 2002, MNRAS, in press

\bibitem[]{} Ho, W. C. G., Lai, D. 2001, MNRAS, 327, 1081

\bibitem[]{} Ibrahim, A. I., et al. 2002a, ApJ, 574, L51

\bibitem[]{} Ibrahim, A. I., Swank, J. H., Parke, W. 2002b, ApJL,
in press

\bibitem[]{} Israel, G.L., et al. 2002, ApJL in press (astro-ph/0209599)

\bibitem[]{} Ruderman, M. A., \& Sutherland, P. G. 1975, ApJ,
196, 51

\bibitem[]{} Mereghetti, S., De Luca, A., Caraveo, P.A., Becker,
W., Mignani, R.,Bignami, G.F. 2002, ApJL, in press
(astro-ph/0207296)

\bibitem[]{} Sanwal, D., Pavlov, G.G., Zavlin, V.E., Teter, M.A.
2002, ApJ, 574, L61


\bibitem[]{} Tr\"umper, J., et al. 1978, ApJ, 219, L105

\bibitem[]{} Wang, Z. X., Chakrabarty, D. 2002, ApJL,
in press (astro-ph/0207540)

\bibitem[]{} Xu, R.X. 2002, ApJ, 570, L65

\end{thebibliography}
\end{document}